\documentclass{article}
\usepackage{spconf,amsmath,graphicx}
\usepackage{booktabs}
\usepackage{multirow}
\usepackage{subfigure}
\usepackage{xcolor}

\title{Exploration of A Self-Supervised Speech Model: A Study on Emotional Corpora}

\name{Yuanchao Li, Yumnah Mohamied, Peter Bell, Catherine Lai}
\address{
  Centre for Speech Technology Research, University of Edinburgh\\
  \{yuanchao.li, ymohamie, peter.bell, c.lai\}@ed.ac.uk}
\makeatletter
\def\ps@IEEEtitlepagestyle{%
\def\@oddfoot{\mycopyrightnotice}%
\def\@evenfoot{}%
}
\def\mycopyrightnotice{%
{\footnotesize 978-1-6654-7189-3/22/\$31.00~\copyright~2023 IEEE\hfill} 
\gdef\mycopyrightnotice{}
}

\copyrightnotice{978-1-6654-7189-3/22/\$31.00~\copyright2023 IEEE}

\begin{document}
%
\maketitle
\begin{abstract}

Self-supervised speech models have grown fast during the past few years and have proven feasible for use in various downstream tasks. Some recent work has started to look at the characteristics of these models, yet many concerns have not been fully addressed. In this work, we conduct a study on emotional corpora to explore a popular self-supervised model -- wav2vec 2.0. Via a set of quantitative analysis, we mainly demonstrate that: 1) wav2vec 2.0 appears to discard paralinguistic information that is less useful for word recognition purposes; 2) for emotion recognition, representations from the middle layer alone perform as well as those derived from layer averaging, while the final layer results in the worst performance in some cases; 3) current self-supervised models may not be the optimal solution for downstream tasks that make use of non-lexical features. Our work provides novel findings that will aid future research in this area and theoretical basis for the use of existing models.
\end{abstract}
\section{Introduction}
\label{sec:intro}

Choosing the right features is a priority in machine learning-based speech tasks. How much target information the features contain fundamentally determines how well a model will work. There are a large number of features to represent and explain the complexity and variability of speech signals in extensive multi-disciplinary studies \cite{fant1973speech,benesty2008springer,pierre2003production}. Task specific features have been widely and effectively used in various speech tasks. For example, cepstral features such as Mel-Frequency Cepstral Coefficients (MFCC), Linear-Frequency Cepstral Coefficients (LFCC), and Perceptual Linear Prediction (PLP) cepstral coefficients dominated Automatic Speech Recognition (ASR) for many years \cite{davis1980comparison, hermansky1990perceptual}. Similarly, other speech tasks have their own preferred feature sets. In Speech Emotion Recognition (SER), surprasegmental features, such as pitch, energy, speaking rate \cite{frick1985communicating, acosta2011achieving}, have proven more helpful than information about phonetic segments. Aspects of speech which are often discarded in automatic transcription, such as disfluencies, are also known to be helpful in tasks such as SER \cite{moore2014word}. The same situation is true for other tasks, such as dialog act detection \cite{fernandez2002dialog}, which leads to handcrafted engineering to understand the contributions of various features.

On the other hand, directly learning feature mappings from speech signals without handcrafted engineering has emerged as a trend during the past decade. Such End-to-End (E2E) approaches benefit from the success of deep learning technologies and have proven useful in many speech tasks, including ASR, SER, speaker verification, and disorder classification \cite{graves2014towards,snyder2016deep,li2019improved,pan2020acoustic}. The E2E approach eliminates the separate step of feature extraction and enables joint training of multiple tasks due to shared representations. This can allow the models to learn feature spaces that are more representative of the actual task than handcrafted features. 

Inspired by the success of Self-Supervised Learning (SSL) in natural language processing \cite{peters-etal-2018-deep,devlin2018bert}, work on addressing the general lack of task-specific labeled speech data has accelerated in the past few years. Most of these approaches can be divided into generative modeling approaches and discriminative modeling approaches \cite{chung2019unsupervised,baevski2020wav2vec,hsu2021hubert}. SSL utilizes information extracted from the input data itself as the label to learn to encode general-purpose representations. These pre-trained upstream models have proven effective for downstream speech tasks, including speaker verification \cite{chen2022does,fan2020exploring} and SER \cite{pepino2021emotion}. However, what these models are actually learning is still understudied and questions and concerns remain about why and how these models benefit downstream tasks: \textit{Are the generated representations optimal for every task? How to utilize them for different purposes?}

With these questions in mind, we study wav2vec 2.0 \cite{baevski2020wav2vec} on emotional corpora, demonstrating how this type of self-supervised model can be explored for downstream tasks. Our experiments show that: 1) wav2vec 2.0 appears to discard some paralinguistic information that is less useful for word recognition purposes and does not treat all emotions and paralinguistic features equally; 2) for SER, representations from the final layer could result in the worst performance in some cases; 3) current self-supervised models need to be carefully fine-tuned to adapt to downstream tasks that make use of non-lexical features. We hope our findings can provide the research community with a new perspective to look at the effectiveness and usage of self-supervised models.

\section{Related Work}
\label{sec:work}

There is no doubt that large-scale speech models using SSL are becoming integral in speech processing tasks. Most of them can be divided into generative or discriminative approaches. The generative approaches generate future frames from past frames, or masked frames from unmasked frames by learning to minimize reconstruction loss \cite{chung2019unsupervised,chung2020vector,liu2020non}.
On the other hand, the discriminative approaches discriminate positive samples from negative samples while minimizing contrastive prediction loss \cite{van2018representation,baevski2020wav2vec,hsu2021hubert}.
These self-supervised models are generally trained on Librispeech \cite{panayotov2015librispeech}, a corpus based on public domain audio books primarily used for ASR research. Although self-supervised objectives are general, the design of popular SSL models has been primarily driven by the goal of improving automatic transcription.

Unlike traditional speech modeling approaches that have been extensively researched, these SSL models have just started to be explored in very recent years, with wav2vec 2.0 (W2V2) attracting the most attention for its wide application potential. For example, Pasad et al. \cite{pasad2021layer} conducted layer-wise analysis of W2V2 using a suite of tools and found 1) acoustic and linguistic properties are encoded in different layers; 2) the pre-trained model follows an autoencoder-style behavior; 3) the model encodes some non-trivial word meaning information. Fan et al. \cite{fan2020exploring} showed that W2V2 has the ability to discriminate between speakers and also languages, and this distinction is more obvious in lower layers. They hence proposed multi-task learning of speaker verification and language identification, and verified its feasibility. Li et al. \cite{li2022fusing} noticed the recognition of longer emotional utterances that contain more contextual information benefits from the contextual characteristic of W2V2. They proposed a joint training scheme by hierarchically fusing multiple W2V2 outputs for SER. Yang et al. \cite{yang2021superb} set up benchmark performance using self-supervised speech models on a range of tasks. 

Nevertheless, these self-supervised speech models are still understudied and the above-mentioned works have limitations. For example, \cite{pasad2021layer} did not extend their exploration to non-ASR downstream tasks. In \cite{fan2020exploring} and \cite{li2022fusing}, only a portion of the layer difference was shown, so misses a thorough layer-wise analysis. In \cite{yang2021superb}, they presented downstream task performance without further explanation. Furthermore, none of those studies investigated paralinguistic characteristics in W2V2 representations. As such, in the current work, we build on previous work while adding new perspectives from detailed quantitative analysis on emotional corpora.

\vspace{-5pt}
\section{Corpora and Model Description}
\subsection{Corpora}


\noindent\textbf{IEMOCAP} (IEM) \cite{busso2008iemocap} has five dyadic sessions with ten actors (five male and five female), each with a scripted and improvised multimodal interaction. The corpus consists of approximately 12 hours of speech that has been annotated by three annotators with ten emotion classes. Following prior research \cite{li2022fusing}, we combined \textit{Happy} and \textit{Excited}, and removed utterances that do not have transcripts, bringing the total number of utterances used in this study to 5,500, each with one label from four classes: \textit{Angry, Happy, Neutral, and Sad}.

\noindent\textbf{RAVDESS} (RAV) \cite{livingstone2018ryerson} contains a speech set and a song set. We only use the speech set, which has 1,440 utterances from 24 actors (12 female, 12 male) in eight emotions: \textit{Calm, Happy, Sad, Angry, Fear, Surprise, Disgust, and Neutral}. Ratings were provided by untrained individuals. In the process of collecting data, the actors spoke two fixed sentences with different classes of emotion, so the corpus has a good balance of emotions. The actors were given two trials for each utterance and asked to produce 60 speech clips in total.


The major reason that we choose to use RAV is that, even though other corpora may have a larger size, it provides fixed sentences with different emotional expressions. Such a setting excludes the lexical influence by ``forcing'' different emotions to have the same linguistic content, thus helping us to better explore the effects of the acoustic properties of W2V2 by eliminating the effects raised by lexical content (e.g., word pronunciation causing prosody variation).

\subsection{Model}
We look at W2V2 \cite{baevski2020wav2vec}, a SSL framework comprised of three major components: a CNN-based local encoder that extracts a sequence of embeddings from raw audio as latent speech representation $Z$, a Transformer network for obtaining context representation $C$, and a quantization module for discretizing $Z$ into $Q$. Following previous work \cite{pasad2021layer}, we focus our attention on the latent representations learned by the Transformer module of W2V2. 
In this work, we use \textit{wav2vec2-base}, \textit{wav2vec2-base-100h}, and \textit{wav2vec2-base-960h} models, which are the pre-trained and fine-tuned models (on 100h and 960h of Librispeech) respectively. We refer to them as \textit{PT}, \textit{FT100}, and \textit{FT960}. We choose W2V2 because it is the most widely used SSL speech model, with the expectation that the exploratory approach can be generalized to similar SSL models.


\vspace{-5pt}
\section{Experiments and Results}
\label{sec:exp}

We perform a set of probing experiments, including the following quantitative measures:

\begin{figure}[t]
  \centering
  \subfigure{\includegraphics[width=0.435\textwidth]{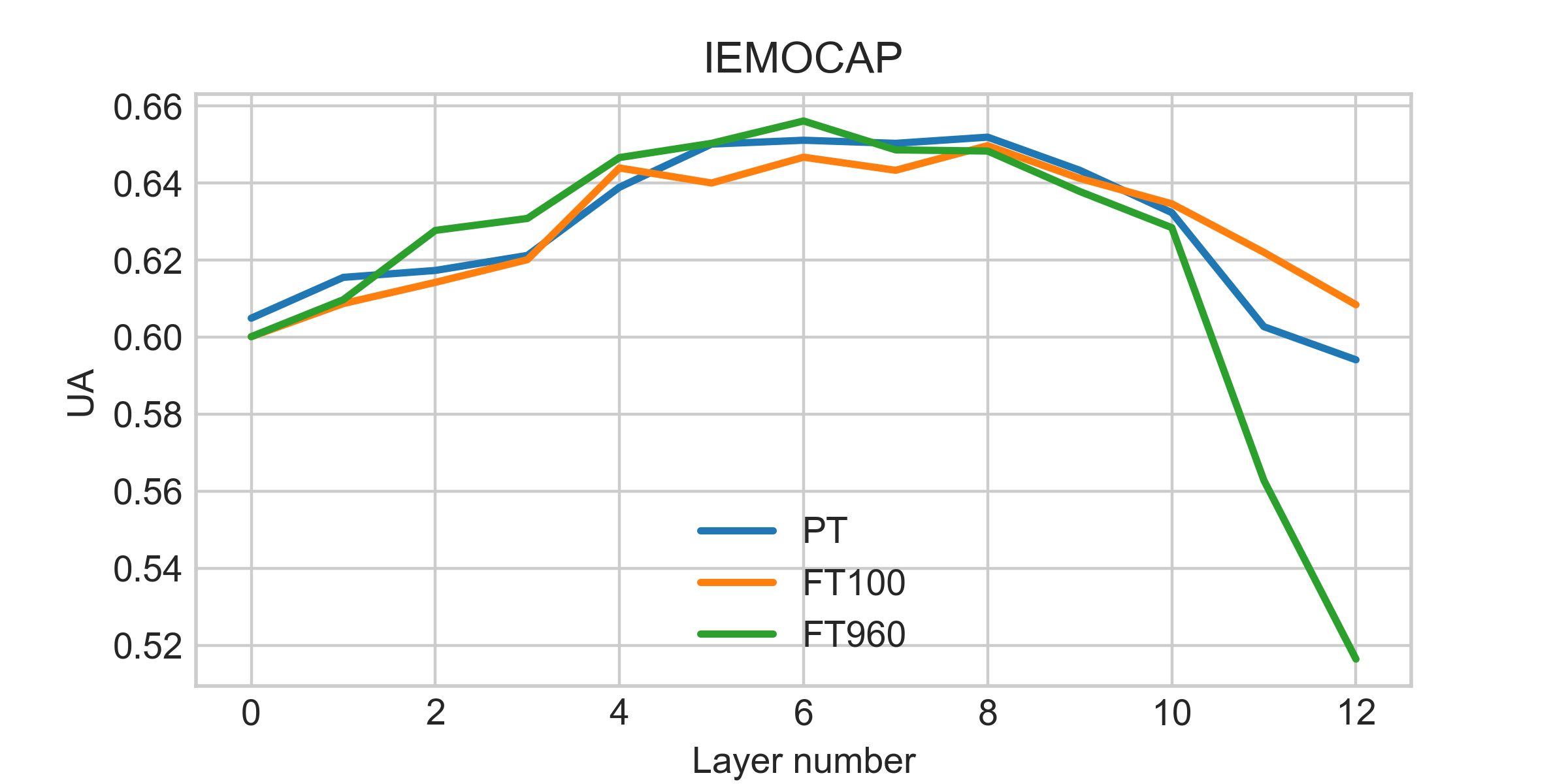}}
  \subfigure{\includegraphics[width=0.435\textwidth]{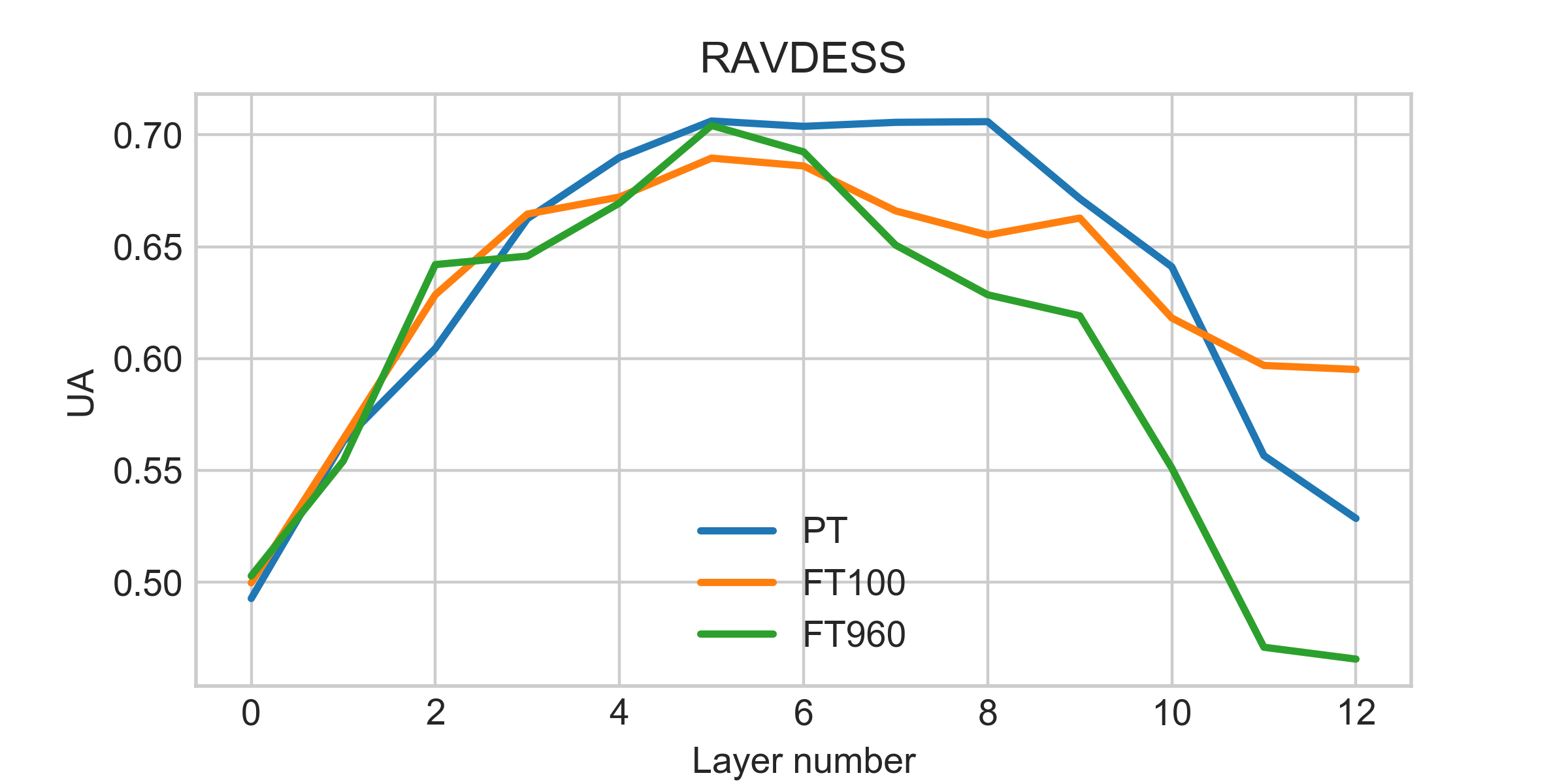}}
  \caption{SER accuracy comparison using different models.}
  \label{fig:acc}
  \vspace{-14pt}
\end{figure}

\noindent\textbf{Probing SER performance}. We first implement a layer-wise analysis by using the output of every individual layer within the Transformer network to demonstrate how information encoded by W2V2 contributes to SER. Next, as there is no common practice of how to utilize W2V2 representations as input features for downstream tasks, we compare the performance of three commonly used approaches of using W2V2 representations as input features: 1) taking the last layer output \cite{chen2021exploring,sharma2022multi,cai2021speech}; 2) taking the average of all layer outputs \cite{boigne2020recognizing}; 3) taking the weighted average of all layer outputs (assigning a trainable weight to each layer output) \cite{pepino2021emotion,yang2021superb}. We also propose a fourth approach which excludes the last two layers from averaging as they generally underperform other layers. We evaluate the performance using Unweighted Accuracy (UA). Like most downstream tasks, we use W2V2 models as frozen feature extractors. 
Since our goal is to explore information in W2V2 representations, we build a simple downstream model comprising only two dense layers (128 and 16 neurons, respectively) with \textit{ReLU} activation and one output layer (four neurons for IEM and eight neurons for RAV) with \textit{Softmax} activation. The learning rate is set as 1e-4 and 2e-4 for IEM and RAV with the \textit{AdamW} optimizer, respectively, and the weight decay is set as 1e-5. The batch size is 64, and we train the models until validation loss converges, as different layer outputs converge at different steps. For IEM, we implement 5-fold cross-validation in accordance with prior works. For RAV, we randomly divide 24 speakers into four groups and implement 4-fold cross validation.

\begin{table}[t]
\centering
\caption{UA (\%) using different inputs and models.}
\label{tab:acc}
\scalebox{0.89}{
\begin{tabular}{llccc}
\hline
\multirow{2}{*}{\textbf{Input}} & \multirow{2}{*}{\textbf{Model}} & \multicolumn{2}{c}{\textbf{Corpus}} \\
 &  & IEM & RAV \\ \hline
\multirow{3}{*}{Best layer} & \textit{PT} & 65.19 & \colorbox{pink}{70.62} \\
 & \textit{FT100} & 64.97 & 68.96 \\
 & \textit{FT960} & \colorbox{lime}{65.61} & 70.42 \\ \hline
 \multirow{3}{*}{Last layer} & \textit{PT} & 59.41 & 52.85 \\
 & \textit{FT100} & \colorbox{lime}{60.84} & \colorbox{pink}{59.51} \\
 & \textit{FT960} & 51.64 & 46.56 \\ \hline
\multirow{3}{*}{Average} & \textit{PT} & 64.93 & 67.26 \\
 & \textit{FT100} & 64.72 & \colorbox{pink}{67.40} \\
 & \textit{FT960} & \colorbox{lime}{65.51} & 64.20 \\ \hline
\multirow{3}{*}{Avg. w/o last two} & \textit{PT} & 65.11 & 67.36 \\
 & \textit{FT100} & 64.90 & \colorbox{pink}{67.50} \\
 & \textit{FT960} & \colorbox{lime}{65.87} & 65.56 \\ \hline
\multirow{3}{*}{Weighted average} & \textit{PT} & 65.28 & 68.47 \\
 & \textit{FT100} & 64.94 & \colorbox{pink}{68.89} \\
 & \textit{FT960} & \colorbox{lime}{65.67} & 65.11 \\ \hline
\end{tabular}
}
\vspace{-14pt}
\end{table}

Fig.~\ref{fig:acc} depicts the trends of layer-wise UA on the two corpora. We include layer 0 (the output of the CNN encoder right before the Transformers) in accordance with prior works. We see that: \textbf{1)} Before the best middle layer (layer 6 for IEM and layer 5 for RAV), all three models (\textit{PT, FT100, and FT960}) show the same trend: accuracies go up and are relatively close, but then start dropping after the middle layer. This upward-downward trend is possibly related to the acoustic-linguistic property of W2V2 \cite{pasad2021layer}. Raw frame-level inputs are encoded by the Transformers until the middle layer. At this point, the representations encode phonetic information but have not yet lost much of the original acoustic properties. This makes the middle layers contain the most useful information for SER. In subsequent layers, the representations gradually encode word identity and word meaning more strongly. At this stage, potential ASR errors with the loss of the original acoustic information have been appeared to lead to drops in SER accuracy. \textbf{2)} On IEM, there are barely any differences among the UAs until layer 11, while on RAV, the differences after the middle layer are more dramatic. This phenomenon is plausible as RAV only has two fixed statements, yet IEM contains various sentences, allowing fine-tuned W2V2 models to make more use of linguistic information, which makes up for the acoustic loss. In RAV, however, every sentence is repeated with each emotion, which means linguistic information has no contribution to emotion discrimination. \textbf{3)} In general, \textit{PT} \textgreater \ \textit{FT100} \textgreater \ \textit{FT960} from the middle layer but \textit{FT100} clearly outperforms the other two on the last two layers. We assume that moderate fine-tuning enables W2V2 to achieve a good acoustic-linguistic balance, as word information has been found encoded by the last two layers in fine-tuned models \cite{pasad2021layer}, and such a balance helps \textit{FT100} achieve better performance on the last two layers.

\begin{figure*}[t]
  \centering
  \includegraphics[width=0.985\textwidth]{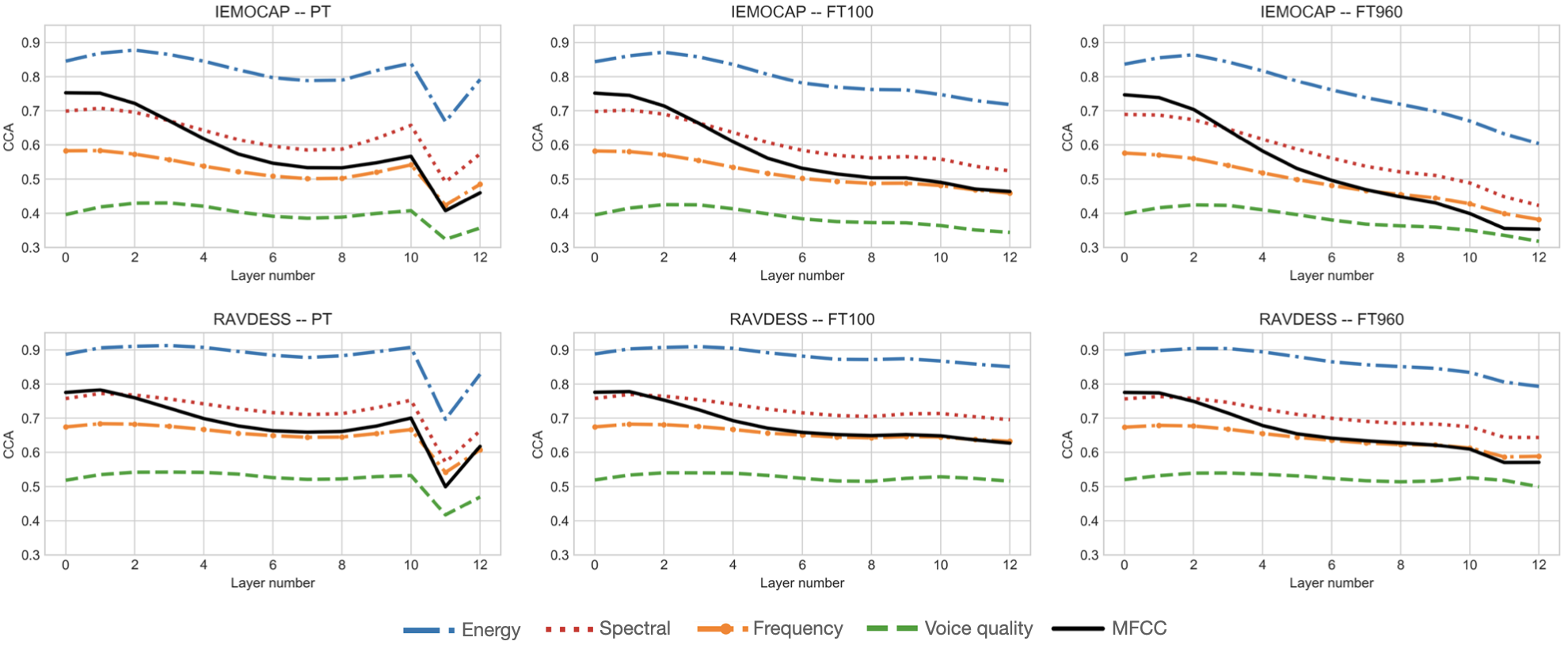}
  \caption{CCA similarity comparison for paralinguistic property.}
  \label{fig:cca}
  \vspace{-15pt}
\end{figure*}

Table~\ref{tab:acc} compares the SER accuracies using different inputs and models, and yields the following findings: \textbf{1)} UAs of the best layer of \textit{PT} and \textit{FT960} are close and higher than \textit{FT100}. So, while moderate fine-tuning enables the model to capture both acoustic and linguistic properties for \textit{FT100}, neither of them is fully encoded causing a decrease in accuracy for the best layer. \textbf{2)} The situation is reversed on the last layer. Compared to \textit{FT100}, \textit{PT} lacks linguistic information and \textit{FT960} relies too much on imperfect linguistic information while losing acoustic information due to ``over'' fine-tuning. \textbf{3)} Word-level information does not help SER on RAV, as mentioned before, which makes the deeper layers of \textit{FT960} the worst. Hence, it is reasonable that \textit{FT960} generates better performance on IEM yet worse performance on RAV by either taking the average on all layers, the average without the last two layers, or the weighted average on all layers as input. \textbf{4)} Corpora like RAV, whose emotions depend only on acoustics, may only need the best middle layer as input for SER. However, it is hard to say which input is obviously the best for corpora like IEM, whose emotions are also affected by text. \textbf{5)} Except for the ``best'' layer inputs, \textit{FT960} and \textit{FT100} produce the best results on IEM and RAV, respectively. This differs from the patterns in Fig.~\ref{fig:acc} from which we would expect \textit{FT960} to perform the worst on IEM and \textit{PT} the best on RAV. It means that the performance obtained by averaging layer outputs does not equal the average of all layer performance, which demonstrates that representations of different layer contain different information contributing to SER. 

\begin{table}[t]
\centering
\caption{Extracted paralinguistic features.}
\label{tab:para}
\scalebox{0.95}{
\begin{tabular}{ll}
\hline
\textbf{Feature set} & \textbf{Low-level descriptors} \\ \hline
Energy & Loudness; Harmonics-to-noise ratio \\ \hline
Frequency & \begin{tabular}[c]{@{}l@{}}Pitch; Formant 1;\\ Formant 1, 2, 3 frequency\end{tabular} \\ \hline
Spectral & \begin{tabular}[c]{@{}l@{}}Alpha ratio; Hammarberg index;\\ Formant 1, 2, and 3 relative energy;\\ Spectral slope 0-500 Hz, 500-1500 Hz;\\ Harmonic difference H1-H2 and H1-A3\end{tabular} \\ \hline
Voice quality & Jitter; Shimmer \\ \hline
\end{tabular}
}
\vspace{-14pt}
\end{table}

\noindent\textbf{Probing paralinguistic information}. 
In this experiment, we measure the similarities between each layer's output and different types of paralinguistic features to see how W2V2 retains well-known acoustic correlates of speech perception. We evaluate the similarity using Canonical Correlation Analysis (CCA) \cite{hardoon2004canonical}. The selected paralinguistic features are listed in Table~\ref{tab:para}, which are mainly based on eGeMAPS \cite{eyben2015geneva}, commonly used as a minimal set of features for SER. We also extract MFCC as linguistic (phone) features for comparison. We downsample them to make their sequence lengths comparable to W2V2 representations as required by CCA.

Fig.~\ref{fig:cca} shows the layer-wise CCA on IEM and RAV using three W2V2 models. \textbf{1)} The curves of paralinguistic features are flatter than those of MFCC, indicating that W2V2 does not focus on as much paralinguistic information. In particular, voice quality features seem barely taken into the encoding process. \textbf{2)} When looking at \textit{PT} models, we can see a reverse trend from the middle layer, which reinforces a view that the \textit{PT} model follows an autoencoder style behavior where deeper layers ``reconstruct'' the input \cite{pasad2021layer}. The reverse trend of similarity with MFCC on IEM is weaker than that on RAV, possibly demonstrating that the linguistic complexity makes the reconstruction process harder and more error-prone. Hence, the deeper representations are more accurate and more similar to MFCC on RAV than on IEM. The peculiar pattern on the last two layers is due to the training objective of masked segment prediction (cf.  \cite{pasad2021layer}). \textbf{3)} When looking at the fine-tuned models, we note that the similarities keep decreasing since the models have been fine-tuned towards ASR and learn to compute speech information from frame to phoneme, and then to word level with layer depth \cite{pasad2021layer}. This phenomenon reinforces our explanation of the accuracy drop in Fig.~\ref{fig:acc} that acoustic properties are being replaced by linguistic ones that contain errors. \textbf{4)} 
The graphs indicate that the overall similarity variation on IEM is larger than on RAV. This is again, likely due to the fact that RAV has much less linguistic variation overall, which we in turn see as less change in CCA through the layers. 
Moreover, since 
how we say a sentence is affected by its content, the CCA variation of paralinguistic features is larger on IEM than on RAV. Finally, since the layer outputs contain more complex linguistic information, the overall CCA values for paralinguistic features on IEM are lower than those on RAV, no matter the starting values or overall values. However, the starting values of similarities with MFCC on both corpora are almost the same, further suggesting that W2V2 focuses on learning linguistic information rather than paralinguistic.

\noindent\textbf{Probing layer correlation}. To better understand how different layer outputs are correlated with each other before and after fine-tuning, and how W2V2 encodes information and contributes to SER, we calculate pair-wise CCA similarities of W2V2 representations from every layer and plot the similarities using heat maps to visualize the correlations. We only discuss IEM, as the same patterns are found on RAV.

\begin{figure}[t]
  \centering
  \includegraphics[width=0.484\textwidth]{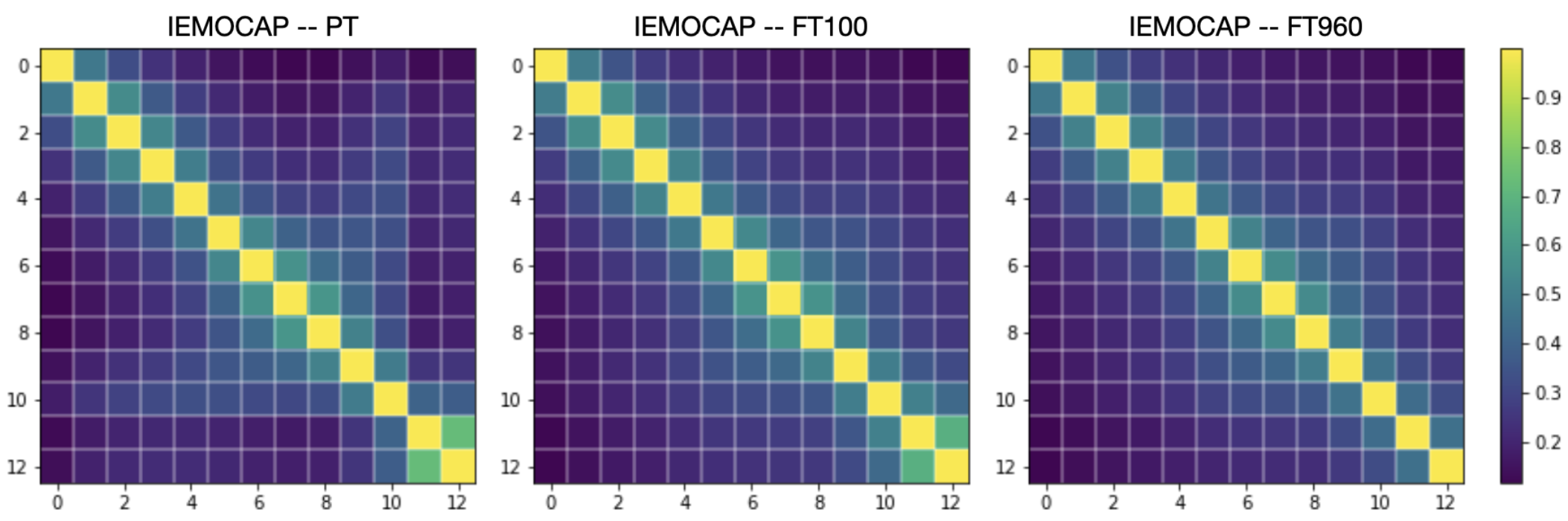}
  \caption{Pair-wise correlations of layer representations.}
  \label{fig:pair}
  \vspace{-15pt}
\end{figure}

\begin{table}[t]
\centering
\caption{Correlations between the last two and prior layers.}
\label{tab:pair}
\scalebox{0.92}{
\begin{tabular}{llccccc}
\hline
\textbf{L.} & \textbf{Model} & \textbf{L. 6} & \textbf{L. 7} & \textbf{L. 8} & \textbf{L. 9} & \textbf{L. 10} \\ \hline
11 & \textit{\begin{tabular}[c]{@{}l@{}}PT\\ FT100\\ FT960\end{tabular}} & \begin{tabular}[c]{@{}c@{}}0.25\\ \colorbox{lime}{0.29}\\ 0.24\end{tabular} & \begin{tabular}[c]{@{}c@{}}0.24\\ \colorbox{lime}{0.30}\\ 0.26\end{tabular} & \begin{tabular}[c]{@{}c@{}}0.24\\ \colorbox{lime}{0.31}\\ 0.27\end{tabular} & \begin{tabular}[c]{@{}c@{}}0.31\\ \colorbox{lime}{0.38}\\ 0.32\end{tabular} & \begin{tabular}[c]{@{}c@{}}0.45\\ \colorbox{lime}{0.53}\\ 0.43\end{tabular} \\ \hline
12 & \textit{\begin{tabular}[c]{@{}l@{}}PT\\ FT100\\ FT960\end{tabular}} & \begin{tabular}[c]{@{}c@{}}0.26\\ \colorbox{lime}{0.27}\\ 0.22\end{tabular} & \begin{tabular}[c]{@{}c@{}}0.25\\ \colorbox{lime}{0.28}\\ 0.22\end{tabular} & \begin{tabular}[c]{@{}c@{}}0.26\\ \colorbox{lime}{0.29}\\ 0.23\end{tabular} & \begin{tabular}[c]{@{}c@{}}0.31\\ \colorbox{lime}{0.34}\\ 0.26\end{tabular} & \begin{tabular}[c]{@{}c@{}}0.43\\ \colorbox{lime}{0.44}\\ 0.32\end{tabular} \\ \hline
\end{tabular}
}
\vspace{-15pt}
\end{table}

Fig.~\ref{fig:pair} shows the pair-wise correlations. \textbf{1)} In \textit{PT} model, we can notice an arrow-like shape from layers 0--10. Specifically, layer 9 and 10 have higher correlations with shallow layers compared to those in \textit{FT100} and \textit{FT960}. This shows that representations start becoming more general in those two layers, indicating pre-trained W2V2 is indeed getting previous information back as assumed by \cite{pasad2021layer}, and also explains the reverse trend in \textit{PT} model. However, such a pattern seems specific to these two layers as it is not obvious in prior deep layers, even layer 8, which also accounts for the reverse trend (Fig.~\ref{fig:cca} in Sec.~\ref{sec:exp} and Fig. 3 in \cite{pasad2021layer}). After fine-tuning, this phenomenon disappears. The two layers become the same as shallow layers that have high correlations with nearby layers, and the correlations get weaker with distance. \textbf{2)} 
In \textit{PT} model, the last two layers are highly correlated with each other, even more obvious than the prior adjacent layers. However, the correlation gets weaker as fine-tuning goes. The color becomes dimmer in \textit{FT100} and further dimmer in \textit{FT960}. Nevertheless, an interesting phenomenon appears in \textit{FT100}: the similarities between last two layers (especially layer 11) and the prior deep layers (layer 6 to 10) become higher. We present the CCA values in Table~\ref{tab:pair}. It is obvious that the correlations in \textit{FT100} are the highest, but they decrease in \textit{FT960}. Moreover, the lower right part is slightly brighter in \textit{FT100} than in \textit{FT960} (values are skipped as limited space) indicating the deeper layers are more correlated with each other, which means low-level linguistic information (e.g., phonetics) generally exists. These phenomena validate our assumption that moderate fine-tuning enables W2V2 to achieve a good acoustic-linguistic balance but over fine-tuning ``forces'' the model to concentrate on learning high-level linguistic properties (e.g., word meaning) towards ASR.


\begin{figure}[t]
  \centering
  \includegraphics[width=0.425\textwidth]{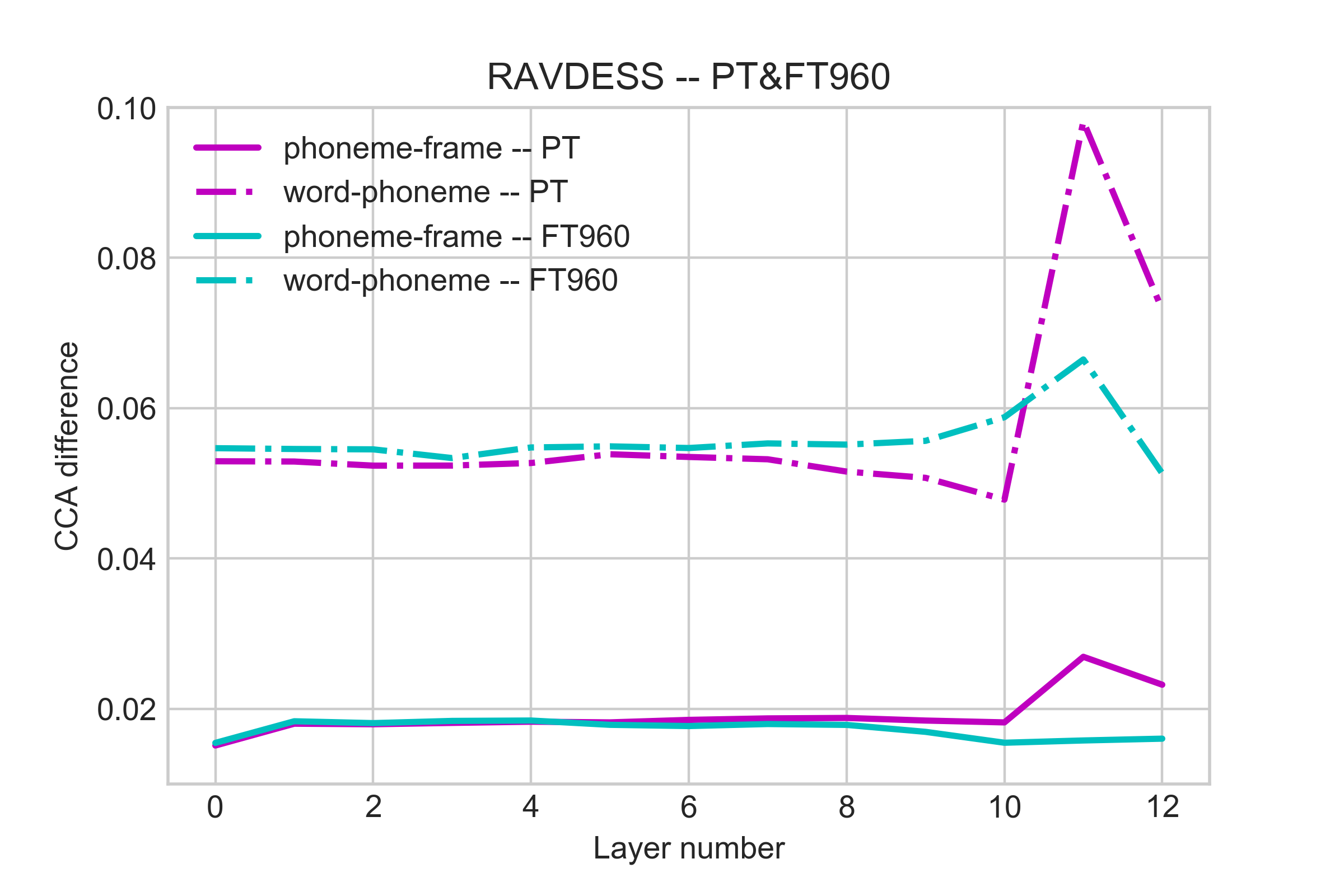}
  \caption{Hierarchical CCA similarity differences.}
  \label{fig:hie}
  \vspace{-15pt}
\end{figure}

\noindent\textbf{Probing hierarchical property}. Since SSL enables frames to capture context information, the representations are expected to contain higher-level meanings. To verify this, we prepare the extracted paralinguistic features at frame, phoneme, and word levels and measure their similarities with W2V2 representations using CCA, respectively. As our purpose is only to verify whether W2V2 features contain high-level speech information, we do not use forced alignment to determine the perfect boundaries. Instead, we adopt a less accurate yet efficient approach to compute hierarchical features: we constitute a phoneme by averaging five consecutive frames (we also tried to add overlap, but the results didn't make much difference): a word by averaging five consecutive phonemes, based on the fact that frame length is set as 25ms when being extracted, and phoneme length varies from 50ms to 200ms (five frames on average) and word length from 250ms to 1,000ms (five phonemes on average) in IEM \cite{chen2022speechformer}. We use all the paralinguistic features provided by eGeMAPS and implement the composition of hierarchical features. Then we downsample the paralinguistic features or the W2V2 representations depending on their lengths to make them comparable. Finally, we compute the CCA differences (${CCA}_{phoneme}-{CCA}_{frame}$ and ${CCA}_{word}-{CCA}_{phoneme}$) which represent how similar the higher-level features with W2V2 representations are compared to the lower-level ones.

From Fig~\ref{fig:hie}, we can note that \textbf{1)} higher-level paralinguistic features do have higher similarities with W2V2 representations as the difference values are all positive. Besides, the value of ${CCA}_{word}-{CCA}_{phoneme}$ is even higher than ${CCA}_{phoneme}-{CCA}_{frame}$, which means W2V2 representations are more similar to word-level paralinguistic information. \textbf{2)} The CCA differences barely change until layer 11 and become larger in the last two layers, which is due to the masked segment prediction enabling them to capture more context information (which is high level), especially on layer 11. \textbf{3)} The curves of fine-tuned models are flatter because the last two layers become more coherent with the other layers by fine-tuning. Note, since the paralinguistic property is affected by linguistics in IEM, the patterns are not as clear as RAV, yet we observed similar trends, so do not discuss it here.

\noindent\textbf{Probing emotion bias}. Different emotions have different paralinguistic patterns \cite{li2007stress,lugger2007relevance}. For example, hot angry and happy emotions usually have high intensity and pitch, while sad and calm emotions have low intensity \cite{luengo2010feature}, Hence, we calculate CCA similarities between paralinguistic features with W2V2 representations of every emotion for discriminative analysis. We also use all the paralinguistic features in eGeMAPS as in the previous task.

\begin{figure}[t]
  \centering
  \subfigure{\includegraphics[width=0.412\textwidth]{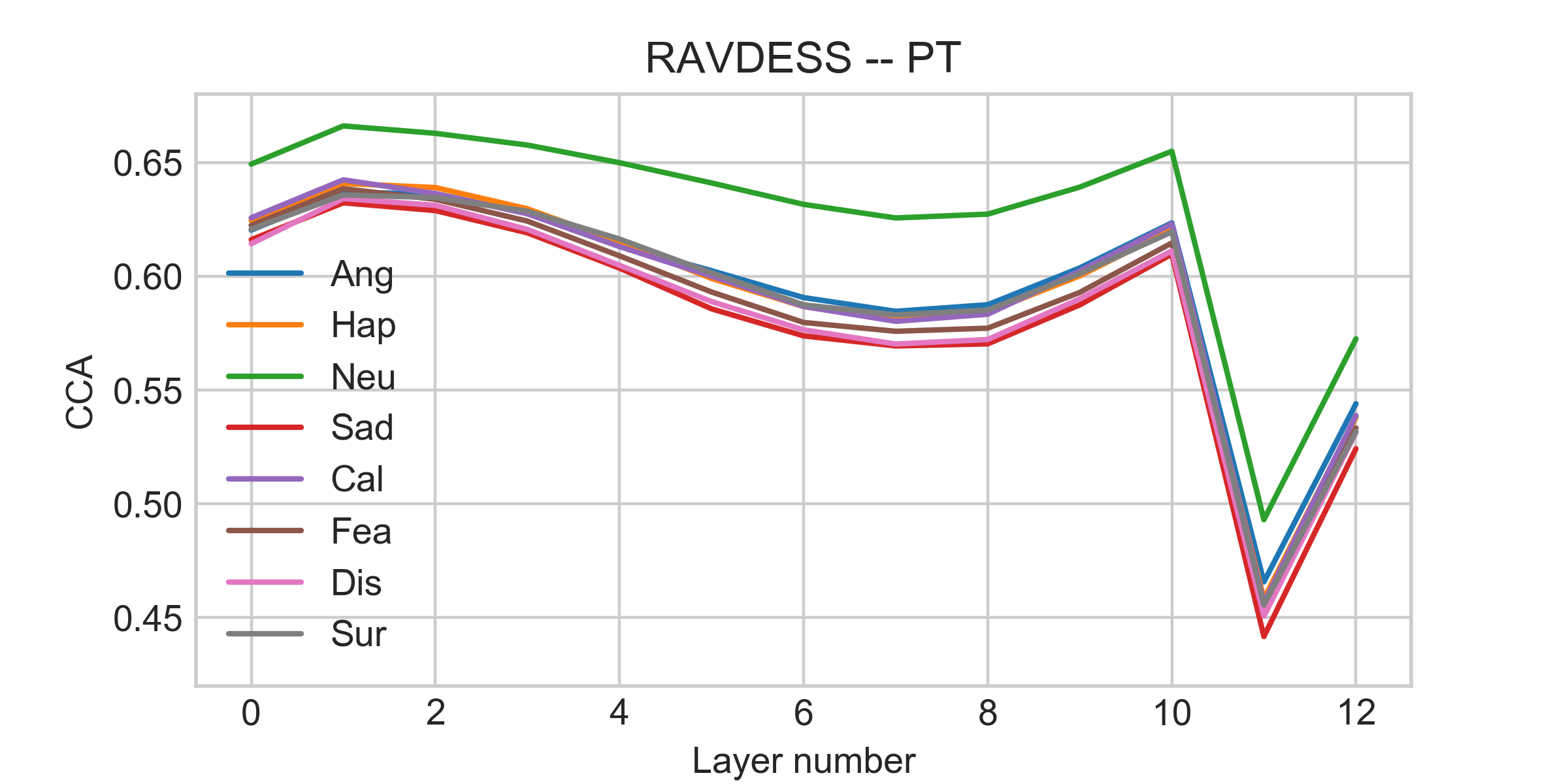}}
  \subfigure{\includegraphics[width=0.412\textwidth]{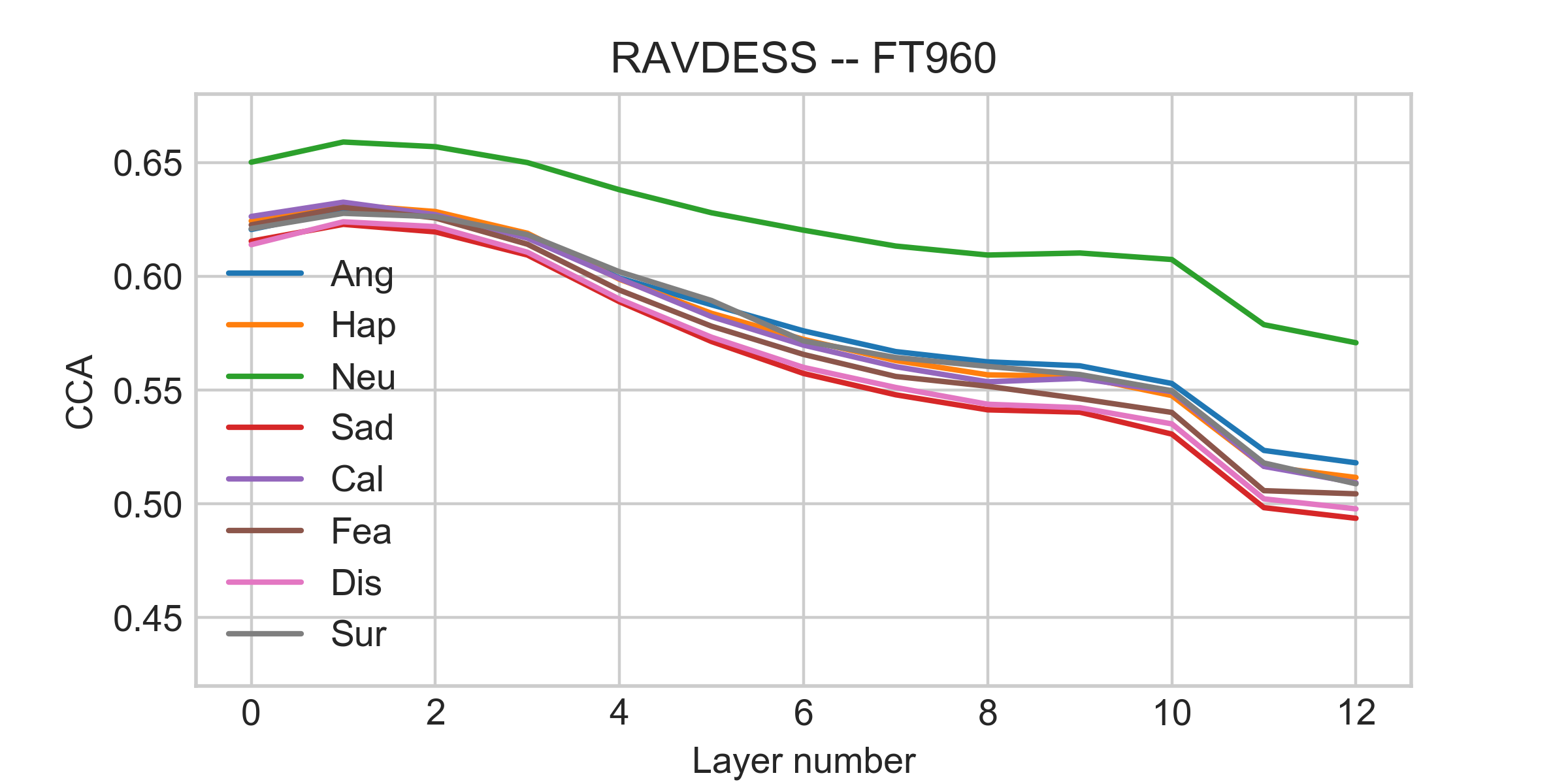}}  
  \caption{Discriminative analysis for emotion bias.}
  \label{fig:dis}
  \vspace{-14pt}
\end{figure}

As illustrated in Fig~\ref{fig:dis}: \textbf{1)} Higher similarities between paralinguistic features with W2V2 representations are found in \textit{Neutral} emotion for both the \textit{PT} model and the \textit{FT} models, pointing to interesting observations: a) \textit{Neutral} is likely more frequently represented within Librispeech, as it is a corpus of read audiobooks where most emotional cues arise only within speech of fictional characters, i.e. bias in the data during pre-training of W2V2 consequently results in learned representations which are emotion-agnostic; b) the pre-training pretext task in W2V2 (predicting masked segments) is not sufficient to learn a truly generalized representation in which different emotions are captured effectively. We also see that the curves converge after the middle layer on \textit{PT} model. This again indicates that the deeper layers (except the last two) of \textit{PT} model reconstruct the acoustic input. \textbf{2)} The curves become even less distinguishable at the last two layers, indicating again the autoencoder type of learning resulting from the masked segment prediction does not help distinguish emotions. This may be because the paralinguistic information of a masked segment is difficult to predict from unmasked segments, as paralinguistic information is more spontaneous and less contextual compared to linguistic information. The masked segment prediction discriminates linguistic information while blurring the paralinguistic difference among frame segments, which makes the paralinguistic properties of every emotion become similar, resulting in the close curves. \textbf{3)} For the \textit{FT960} model, the distances between the curves increase with depth. It seems that W2V2 not only avoids encoding paralinguistic information, but consistently discards some paralinguistic features as learning proceeds, especially in speech that contains rich paralinguistic information, e.g., voice quality, which leads to increasing differences in the similarities between W2V2 representations and paralinguistic features across emotions (otherwise, the distances should not change with layer depth).


\vspace{-5pt}
\section{Discussion}
\label{sec:dis}


1) Fine-tuning affects W2V2 by transforming it from an acoustic-aware model into a linguistic-aware model. Layers of the first half of the Transformer are responsible for encoding acoustic information, as all three models show almost the same patterns. The latter half starts encoding linguistic information as pattern differences occur, but an exception is that the last two layers of \textit{PT} model reconstruct the input.

2) W2V2 should be used with caution on downstream tasks because it potentially loses important paralinguistic information. As information that is not helpful to ASR is discarded with layer depth on fine-tuned models, \textit{PT} model is a better choice for tasks that are largely paralinguistic-dependent. Moreover, the best layer outperforms layer averaging for SER, while the last layer could be the worst choice.

3) While W2V2 (and possibly other similar SSL models) is a universal solution, it is not suitable for all downstream tasks. It does not, for example, outperform previous SER works that take raw signals as input but use less sophisticated end-to-end structures \cite{li2019improved}. Besides, as some paralinguistic information is largely involved in pragmatics such as turn-taking and backchanneling \cite{ward2016interactional}, the deep layers of W2V2 may not be able to model these dialog-level functions.

\vspace{-5pt}
\section{Conclusions}
\label{sec:con}

In this work, we study W2V2 by conducting a set of quantitative 
analysis on emotional corpora. We found W2V2 lacks the ability to capture paralinguistic information. We also contribute to understanding the types of representations W2V2 learns by thoroughly comparing layer outputs in their correlations and SER. The hierarchy and emotion analysis pave the path for better usage of W2V2 on downstream tasks. Our results confirm the assumptions of previous work and strengthen their conclusions, as well as provide novel findings towards better leveraging of SSL speech models.

\vspace{-5pt}
\section{ACKNOWLEDGMENTS}
\label{sec:ack}

We would like to thank the authors of \cite{pasad2021layer} for their kind help and valuable discussion.

\bibliographystyle{IEEEbib}
\bibliography{strings,refs}

\end{document}